\documentclass[a4paper,fleqn]{llncs}

\usepackage{array}
\usepackage{amssymb}
\usepackage{graphicx}
\usepackage{amsmath}
\usepackage{float}
\usepackage{url}
\usepackage{algorithm}
\usepackage{algorithmic}
\urldef{\mailsa}\path|{alfred.hofmann, ursula.barth, ingrid.haas, frank.holzwarth,|
\urldef{\mailsb}\path|anna.kramer, leonie.kunz, christine.reiss, nicole.sator,|
\urldef{\mailsc}\path|erika.siebert-cole, peter.strasser, lncs}@springer.com|    
\newcommand{\keywords}[1]{\par\addvspace\baselineskip
\noindent\keywordname\enspace\ignorespaces#1}

\begin{document}

\mainmatter  

\title{A Statistical Explanation of the Timing Attack on QC-MDPC Code Crypto-system}


%
%

\author{Han Li}%

%


\institute{University of California, Davis
}

%
%

\toctitle{Lecture Notes in Computer Science}
\tocauthor{Authors' Instructions}
\maketitle
\pagestyle{plain}

\begin{abstract}
The McEliece cryptosystem based on quasi-cyclic moderate-density parity-check (QC-MDPC) codes is first purposed in 2013\cite{QCMDPC} and is considered a promising contender in the post-quantum era. Understanding its security is hence essential. Till now, the most effective attacks are the reaction attack\cite{Reaction} and the timing attack\cite{Timing}. Both of these attacks rely on the decoding performance to recover the private key. The reaction attack relies on the decoding failure rate and the timing attack relies on the iterations during decoding. However, the mechanics behind these attacks remain elusive. In this paper, a mathematical model is proposed to explain both attacks by connecting the spectrum of private key and first-layer performance of the decoder.
\end{abstract}

\section{Introduction}

Code-based cryptography is first proposed in 1978 by McEliece\cite{McEliece}. In the original scheme, McEliece exploits Goppa code to store and protect data. The original scheme remains unbroken in the last forty years and becomes one of the most promising crypto-systems that can survive under quantum attacks. However, the original scheme suffers from a large key size. The public key size is the modified generator matrix of the Goppa code we use. Thus, to transfer a length $k$ message, we will need a public key size $k\times n$.

In 2013, a new variant of the McEliece system based on QC-MDPC(quasi-cyclic moderate-density-parity-check code) was proposed\cite{QCMDPC}. This scheme exploits QC-MDPC code, instead of Goppa code, to store data. The quasi-cyclic property can considerably shorten public key size for a given security level, from $\mathcal{O}(n\times k)$ to $\mathcal{O}(n)$. Moreover, the MDPC property makes this code an inferior version of the LDPC code. Therefore, decoding algorithms of LDPC code, usually the Gallagher algorithm, can be directly applied to the QC-MDPC code.

However, this new scheme brings new security flaws. In 2016, Guo Qian\cite{Reaction} proposed a reaction attack on this scheme. In 2018, Edward Eaton proposed a timing attack\cite{Timing}. Both attacks are based on the feedback of the decoding performance during decryption: Reaction attack is based on failure feedback and timing attack is based on the number of iterations during decoding through a side-channel attack. By collecting a large number of data about the decoding performance, we are able to recover the private key with a large probability.

The key idea of the Reaction attack\cite{Reaction} and Timing attack\cite{Timing} is that the private key can be efficiently recovered from the spectrum of the private key. Moreover, we find that the reaction performance and timing performance are influenced by the "overlap" between the spectrum of the private key and the spectrum of error. Then, with the knowledge of performance and error, we can recover the spectrum of the private key and, thus the private key.
\subsection{Our Contribution}

In this paper, we analyze the mechanics behind the two attacks, especially the timing attack. We at first explain why the two attacks work from a coding theory perspective. Then, a mathematical framework of these two attacks is built that can help us understand or estimate the speed of similar attacks.

In sections 3 and 4, we explain why the "overlap" between the spectrum of the private key and the spectrum of error can make difference to the performance of the decoder through connecting the overlapping spectrum to the first layer performance in the decoder. Moreover, we build a series of linear models to describe the relationship numerically.

In section 5, we build a math model to explain how to apply the conclusion in section 3 or 4 to design an attack on the QC-MDPC code crypto-system. Also, this model can help to estimate the speed of such attack.

\section{ QC-MDPC McEliece System and Attacks}

This section will introduce some basic knowledge. Subsection one will focus on the QC-MDPC code, its decoding algorithm, and application to McEliece system. Subsection two will introduce the reaction attack and timing attack on QC-MDPC crypto-system.

\subsection{QC-MDPC Crypto-system}

The QC-MDPC PQC is defined by four parameters $[n,k,\omega,t]$, which represents a class of QC-MDPC codes. The encryption/decryption of QC-MDPC PQC is based on the encoding/decoding process of this code. 

\subsubsection{Code Generation}

The first stage is to generate a QC-MDPC code with the given parameters. In most practical cases, we assume $n=2k$. The process is as following:
\begin{itemize}
    \item Generate two vector $h_0,h_1\in \mathbb{F}_2^k$, each vector is sparse with approximately $\omega/2$ non-zero term(1's).
    \item Generate two $k\times k$ matrix $H_0,H_1$ by right shifting the vectors $h_0,h_1$.For example, given $h_0=[h_0^1,h_0^2,h_0^3,...,h_0^k]$, the corresponding $H_0$ should be:
    \item Define $H=[H_0|H_1]$. This $H$ is the parity-check matrix of our code. All code-word $c\in \mathbb{F}_2^n$ belongs to the null space of $H$ as $cH^T=0$. 
    \item Compute the systematic generator matrix $G=[I_k|Q]$, where $I_k$ is the identity matrix of dimension $k$ and $Q=(H_{1}^{-1}H_0)^T$. This $G$ represents the null space of $H$ as $mGH^T=0$ for all message $m\in \mathbb{F}^k_2$.
\end{itemize}

\subsubsection{QC-MDPC Crypto-system encryption and decryption}

Bob wants to communicate with Alice with the QC-MDPC cypto-system. He will first generate the a $[n,k,\omega,t]$ code $(G,H)$ based on the above method. Then, he will publicly announce $[n,k,t]$ and $G$.

Alice will divide her messages into $k$-bit pieces $m$. The encryption process is:
\begin{itemize}
    \item Generate an error vector $e\in \mathbb{F}^n_2$ with exactly $t$ non-zero term.
    \item Ciphertext $c=mG+e$.
\end{itemize}

Then, when Bob received the ciphertext, he will process the decryption process:
\begin{itemize}
    \item Call the decoder of this QC-MDPC code, $\mathcal{D}_H$ on ciphertext $c$. The decoder will return an unpolluted code-word $c'$ or a symbol of decoding failure.
    \item If the decoding succeeds, Bob can recover the message $m$ simply by cutting the first $k$-bits of $c'$.
    \item If the decoding failure, Bob may require a re-transmission or process other functions.
\end{itemize}

The decoder may succeed but return an wrong message to Bob. QC-MDPC code can reduce the probability of returning wrong message by lowering the value of $\omega$ and $t$. However, lowering $\omega$ and $t$ will reduce the security of this cryptosystem at the same time. We need a trade-off between the performance of codes and performance of cryto-system. The purposed parameters for different security level\cite{QCMDPC}:
\begin{table}[H]
    \centering
    \begin{center}
    \begin{tabular}{| m{5em} | m{1cm}| m{1cm} | m{1cm}|m{1cm}|}
    \hline\hline
        Security Level & $n$ & $k$ & $\omega$ & $t$\\
    \hline\hline
        80 & 9602 & 4801 & 90 & 84 \\
        128 & 20326 & 10163 & 142 & 134 \\
        256 & 65542 & 32771 & 274 & 264 \\
        \hline
    \end{tabular}   
    \end{center}

    \caption{A table to show the purposed parameters under different security levels}
    \label{tab:my_label}
\end{table}

\subsubsection{Decoding algorithm for QC-MDPC code}

The QC-MDPC code has a similar structure to the LDPC(low-density-parity-check) code. Therefore, we can adopt the decoding algorithm of LDPC codes.

The most widely-used algorithm for decoding LDPC codes is the bit-fliping algorithm. This algorithm will require a vector of threshold $b\in \mathbb{Z}^n$ and the parity-check matrix $H$. This algorithm is an iterative algorithm. In each iteration, the algorithm first computes the syndrome(parity nodes) of
the codeword, $s=H*c$ in $\mathbb{F}_2$ field. Then, the algorithm will give each bit a "score" based on $s$ and flip all bits whose score is large than the threshold. If the syndrome $s$ becomes all zeros before the max turn, the decoding is successful; Otherwise, the decoding fails. The pseudo-code for this algorithms is at Appendix A.

\subsection{Timing attack for QC-MDPC systems}

The parity-check matrix $H$ can be fully constructed from its first row $h=[h_0|h_1]$. Moreover, with the knowledge of generator matrix $G$, attackers can recover $h_1$ from $h_0$. So, an attack on $h_0$ is sufficient to break this system.

At first, I introduce the concept of the \textit{distance spectrum} of vectors on $\mathbb{F}_2$.
\subsubsection{Distance between two nonzero term and Distance spectrum of a vector} Given a vector $v\in\mathbb{F}_2^n$ where $v_i,v_j=1,j>i$. The distance between $v_i$ and $v_j$ is:
$$dist(v_i,v_j)=min(j-i,\left\lfloor \frac{n}{2} \right\rfloor-j+i)$$
Distance spectrum $\Delta(v)$ of the vector $v\in\mathbb{F}_2^n$ is a vector in $\mathbb{Z}$ with length $\left\lfloor \frac{n}{2}\right\rfloor$ that contain all the distances of $v$. In detail, $\Delta(v)_k=r$ if and only if there exists exactly $r$ distinct pairs of index $(i_1,j_1),(i_2,j_2),...,(i_r,j_r)$ such that:
$$dist(v_i,v_j)=k$$

By Guo's paper\cite{Reaction}, we have an efficient algorithm to recover the secrete key $h_0$ from its spectrum $\Delta(h_0)$. A timing attack will recover the spectrum of $h_0$ by estimating the average iterations needed to decode codewords of a certain type. The detailed timing attack algorithm is shown in Appendix A.

This algorithm will return a ratio $iteration_i/observed_i$, which converges to $E[iteration|\Delta(e)_i\neq0]$. The timing attack is totally based on an observation that This $E[iteration|\Delta(e)_i\neq0]$ has a negative relationship with the spectrum of secret key $\Delta(h_0)_i$. In the next two section, I will use a linear model to describe this relationship and analyze how this result leads us to a successful attack.

\section{Causal Relation between the Error and the First-layer Performance}

The timing attack relies on the different performance of the decoder to reveal the secrete key. Therefore, an understanding of how the decoder reacts to the different errors can help us analyze the attack. To formulate the problem, consider a QC-MDPC code with the parity-check matrix $H=[H_0|H_1]$, where $H_0,H_1\in \mathbb{F}_2^{k\times k}$ are cyclic. I denote the $h_0$, $h_1$ the first rows of $H_0$, $H_1$ and $\omega_0$, $\omega_1$ the weight of $h_0$ and $h_1$. In practice, $\omega_0$ should be close to $\omega_1$. Otherwise, this code will be too bad to use. An error $e=[e_0|e_1]$ with weight $t$ is sent to the decoder of this code.

I define $\theta$ a measure of "overlaping level" between the spectrum of $h_0$ and the spectrum of $e$.
\begin{equation}
  \theta = \Delta(h_0)*\Delta(e)
\end{equation}

To characterize the first-layer performance of the decoding of error $e$, I count two quantities: ErrGen and ErrCrt. ErrGen is the number of error-less bits that mistakenly flipped in the first iteration, and ErrCrt is the number of erroneous bits that successfully corrected in the first layer. I will show how these two quantity react to $\theta$ and influence the final performance of the decoder.

\subsection{From Error to ErrCrt}

In the first layer, the decoder will flip all the bits connecting to more than the threshold number of unsatisfied parity-check nodes(the score of this bit). So, ErrCrt is largely determined by the distribution of score of erroneous bits, especially the expectation. The relationship between $\theta$ and ErrCrt can be derived from the relation between $\theta$ and the expectation of score of an erroneous bit, $Score_1$.

I assume a linear model between $\theta$ and $Score_1$:
\begin{equation}
  Score_1=\beta*\theta+b+\epsilon
\end{equation}

In practice, we do not care much about the value of $b$. The value of $\epsilon$ is generally estimated in experiment. The value of $\beta$ represents the strength of relation and can be approximate in the following way.

\subsubsection{Without any information on errors.}
 
Error occurs with uniform probability on the first half part of codeword. So:

\begin{equation}
\begin{aligned}\label{eq:pareto mle2}
     E[Score_1]=E[Score(c_i)|e_i=1]=\sum^{\omega_0}_{j=1}E[\mathbf{1}^j_i|e_i=1]=\omega_0\sum^{\omega_0-1}_{\mathrm{m\ is\ even}}\frac{{t-1 \choose m}{k-t \choose \omega_0-1-m}}{{k-1 \choose \omega_0-1}}
\end{aligned}
\end{equation}

\begin{figure}[H]
    \centering
    \includegraphics[width=8.0cm]{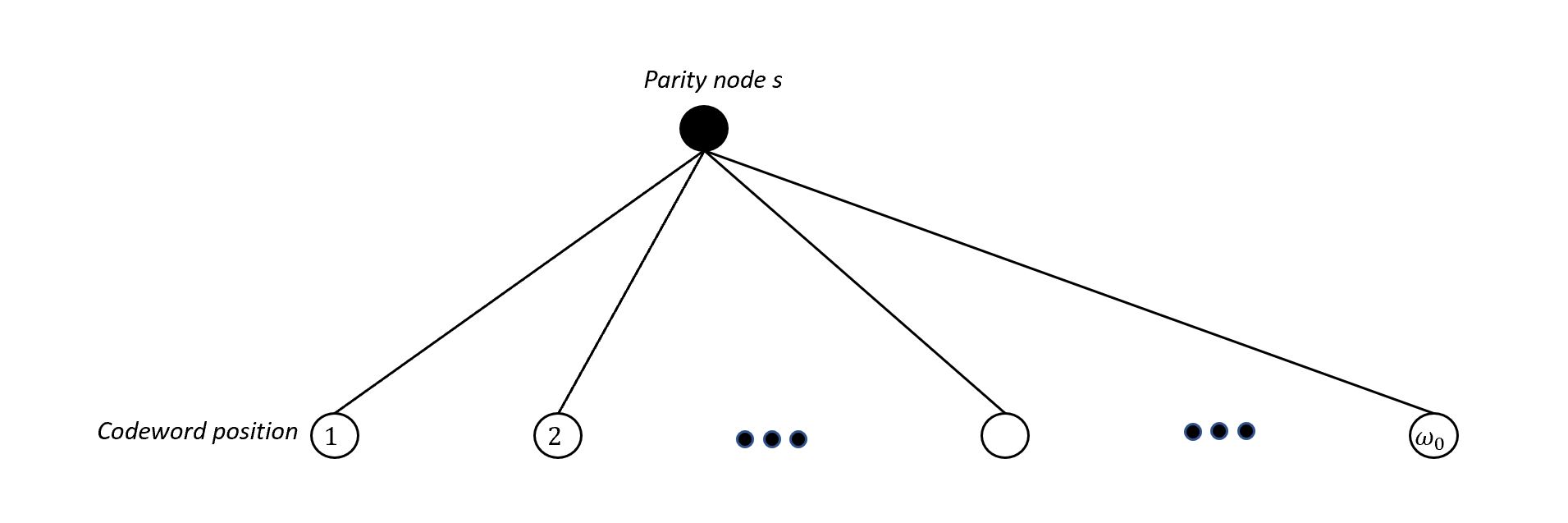}
    \caption{A figure to the relative position between parity nodes and codewords}
    \label{fig:my_label}
\end{figure}

The notation $\mathbf{1}^j_i$ means the indicator function for event that bit node $i$ sits on position $j$ of a parity node $s$, and this $s$ equals 1.

\subsubsection{Event $A$: Given that spectrum of error share a distance d with the spectrum of $h_0$ }. That's, for a distance $d$ such that $\Delta(h_0)_d\not=0$, we know that $\Delta(e)_d$ also increase 1. Let $\{a_i\},i\leq\omega_0$ be the index of non-zero term of $h_0$. Without loss of generality, we assume that $a_2-a_1=d$.
\begin{equation}
\begin{aligned}\label{eq:pareto mle2}
    E[Score_1|A]&=E[Socre(c_i)|e_i=1,A]=\frac{1}{t}\sum^t_{i=1}E[Score(c_{a_i})]\\&=\frac{1}{t}\sum^t_i\sum^{\omega_0}_{j=1}E[\mathbf{1}^j_{a_i}]=\sum^{\omega_0}_{j=1}\frac{1}{t}\sum^{t}_{i}E[\mathbf{1}^j_{a_i}].
\end{aligned}
\end{equation}

Then, we can compute this expectation by elaborating  $\frac{1}{t}\sum^t_{i=1}E[\mathbf{1}_{a_i}^j]$ for different $j$.This expectation can be calculated under two cases.

\begin{itemize}
    \item Case1: $j=1$ or $j=2$:We already know the distance between position 1 and 2 matches the distance between $a_1$ and $a_2$. So, if $a_1$ fall in position 1 or $a_2$ fall in position 2 (each with probability $\frac{1}{t})$, we immediately know $a_2$ fall in position 2 or $a_1$ fall in position 1, respectively. Therefore, the event $\mathbf{1}_{a_1}^1$ and event $\mathbf{1}_{a_2}^2$ is equivalent to that randomly chooses $\omega_0-2$ from $k-t$ zeros and $t-2$ ones without replacement, and gets odd number of ones. By symmetry, the distribution for $j=1$ or $j=2$ is identical.
    \begin{equation}
        \begin{aligned}\label{eq:pareto mle2}
    E[\mathbf{1}_{a_1}^1]=E[\mathbf{1}_{a_2}^2]=\sum_{\mathrm{m\ is\ odd}}^{\omega_0-2}\frac{{t-1\choose m}*{k-t \choose \omega_0-2-m}}{{k-2 \choose \omega_0-2}}
        \end{aligned}
    \end{equation}
    If $a_1$ does not fall in position 1 or $a_2$ does not fall in position 2, the knowledge of event $A$ will not give us any useful information. So, the expectation should not differ from the general cases:
    \begin{equation}
        \begin{aligned}\label{eq:pareto mle2}
    E[\mathbf{1}_{a_1}^1|i\not=1]=E[\mathbf{1}^2_{a_i}|i\not=2]=\sum^{\omega_0-1}_{\mathrm{m\ is\ even}}\frac{{t-1\choose m}*{k-t \choose \omega_0-1-m}}{{k-1 \choose \omega_0-1}}
        \end{aligned}
    \end{equation}
    Together, the expectation of Case 1 is:
    \begin{equation}
        \begin{aligned}\label{eq:pareto mle2}
            \frac{1}{t}E[\mathbf{1}_{a_i}^1|i=j]+\frac{t-1}{t}E[\mathbf{1}_{a_i}^1|i\not=j]
        \end{aligned}
    \end{equation}
    
    \item Case2: $j\not=1,2$
    We need to discuss the value of position 1 and 2 first. The detail process is given in the following figure. The notation $Hyper(N,K,n,even/odd)$ represents the summation of hyper-geometric distribution with parameters $[N,K,n]$ and all even/odd $k$.
    
    \begin{figure}[H]
    \centering
    \includegraphics[width=4.0cm]{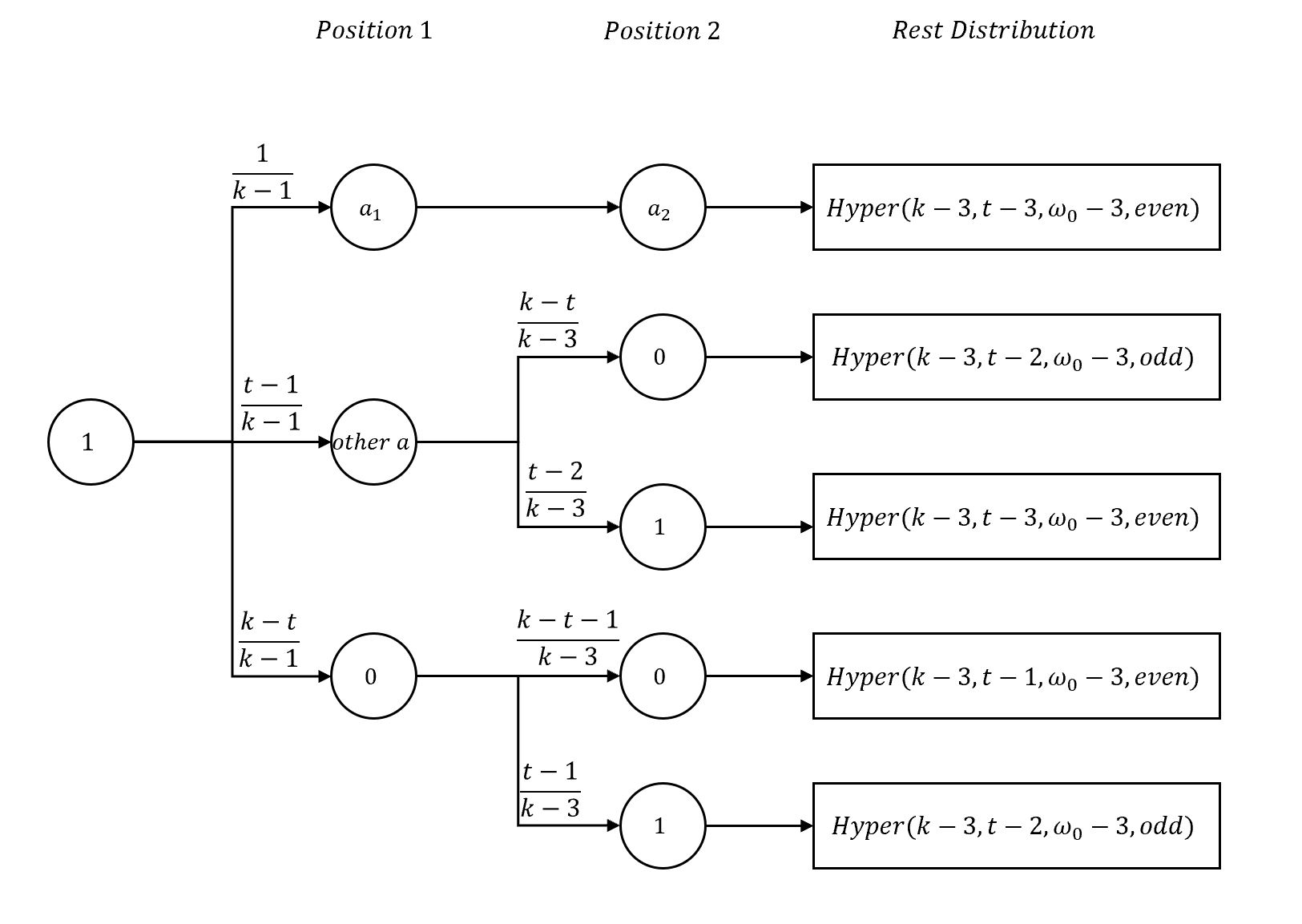}
    \caption{The process to compute the unsatisfied probability of parity-node under case2}
    \label{fig:my_label}
\end{figure}
    
\end{itemize}

This process will give us the expectation under case2 is:
\begin{equation}
\begin{aligned}\label{eq:pareto mle2}
    &\frac{1}{k-1}\sum^{\omega_0-3}_{\mathrm{m\ is\ even}}\frac{{t-3\choose m}{k-t \choose \omega_0-3-m}}{{k-3 \choose \omega_0-3}}+\\&\frac{t-1}{k-1}(\frac{k-t}{k-3}\sum^{\omega_0-3}_{\mathrm{m\ is\ odd}}\frac{{t-2\choose m}{k-t-1 \choose \omega_0-3-m}}{{k-3 \choose \omega_0-3}}+\frac{t-2}{k-3}\sum^{\omega_0-3}_{\mathrm{m\ is\ even}}\frac{{t-3\choose m}{k-t \choose \omega_0-3-m}}{{k-3 \choose \omega_0-3}})+\\&\frac{k-t}{k-1}(\frac{k-t-1}{k-3}\sum^{\omega_0-3}_{\mathrm{m\ is\ even}}\frac{{t-1\choose m}{k-t-2 \choose \omega_0-3-m}}{{k-3 \choose \omega_0-3}}+\frac{t-1}{k-3}\sum^{\omega_0-3}_{\mathrm{m\ is\ odd}}\frac{{t-2\choose m}{k-t-1 \choose \omega_0-3-m}}{{k-3 \choose \omega_0-3}})
\end{aligned}
\end{equation}

From above analysis and my linear model, $E[Score_1]=\beta_1*E[\theta]+b_1$ and $E[Score_1|A]=\beta_1*(E[\theta|A])+b=\beta_1*(E[\theta]+1)+b_1$. So, the difference $E[Score_1|A]-E[Score_1]$ will give us an estimate of the slope around the expectation.

\subsection{From Error to ErrGen}

This case is more complex than the erroneous case. We want to calculate $E[Counter(c_i )|c_i=0]$. However, in my analysis, all erroneous bits appear in the first half part. But during decoding, decoder may flip bits in the second half part of codewords. Luckily, bits in the second half part reacts similarly compared to bits in the first half so that we can apply our analysis of the first half part directly to the second half. The detailed proof is in Appendix C.

\begin{equation}
        \begin{aligned}\label{eq:pareto mle2}
            E[Score_0]=\frac{k-t}{2k-t}E_{first}[Score_0]+\frac{k}{2k-t}E_{second}[Score_0]
        \end{aligned}
\end{equation}

\subsubsection{Without any information}

Without any information on the secrete key and error pattern, we can only regard them as uniformly random vectors each with weight $ω_0$ and $t$. The indicator event $\mathbf{1}^j_i$ under this case is equivalent to that randomly chooses $ω_0-1$ from $k-t-1$ zeros and $t$ ones without replacement, and gets odd number of ones, which follows a hyper-geometric distribution.
\begin{equation}
        \begin{aligned}\label{eq:pareto mle2}
            E_{first}[Score_0]=\sum^{\omega_0}_{j=1}E_{first}[\mathbf{1}^j_i|e_i=0]=\omega_0\sum^{\omega_0-1}_{\mathrm{m\ is\ odd}}\frac{{t\choose m}*{k-t-1 \choose \omega_0-1-m}}{{k-1 \choose \omega_0-1}}
        \end{aligned}
\end{equation}

\subsubsection{Event A: spectrum of error and spectrum of $h_0$ share a distance $d$}
 I assume the distance between position 1 and position 2 is $d$ and $\{a_i,i=1,…,t\}$ be the index of each bit node that equals one, arranging from smallest to largest, where $a_2-a_1=d$.

\begin{equation}
        \begin{aligned}\label{eq:pareto mle2}
            E_{first}[Score_0|A]=\frac{1}{k-t}\sum^{k-t}_{i=1}E_{first}[Score(e_i),i\not \in {a}]=\\\frac{1}{k-t}\sum^{k-t}_i\sum^{\omega_0}_{j=1}E_{first}[\mathbf{1}^j_i,i\not\in{a}]=\sum^{\omega_0}_{j=1}\frac{1}{k-t}\sum^{k-t}_i E_{first}[\mathbf{1}^j_i,i\not \in {a}]
        \end{aligned}
\end{equation}

\begin{itemize}
    \item Case1: $j=1$ and $j=2$: if $j=1$, the position 2 cannot fall to $a_2$ and thus has less probability to be 1. We need discuss the value of position 2 first.
    
    If position 2 connected to 1 with probability $\frac{t-1}{k-2}$: The rest will be identical to the hyper-geometric distribution that draws $ω_0-2$ from $t-1$ ones and $k-t-1$ zeros and sums about even ones.
    
    If position 2 connected to 0 with probability $\frac{k-t-1}{k-2}$:The rest will be identical to the hyper-geometric distribution that draws $ω_0-2$ from $t$ ones and $k-t-2$ zeros and sums about odd ones.
    
    \begin{figure}
        \centering
        \includegraphics[height=4.0cm]{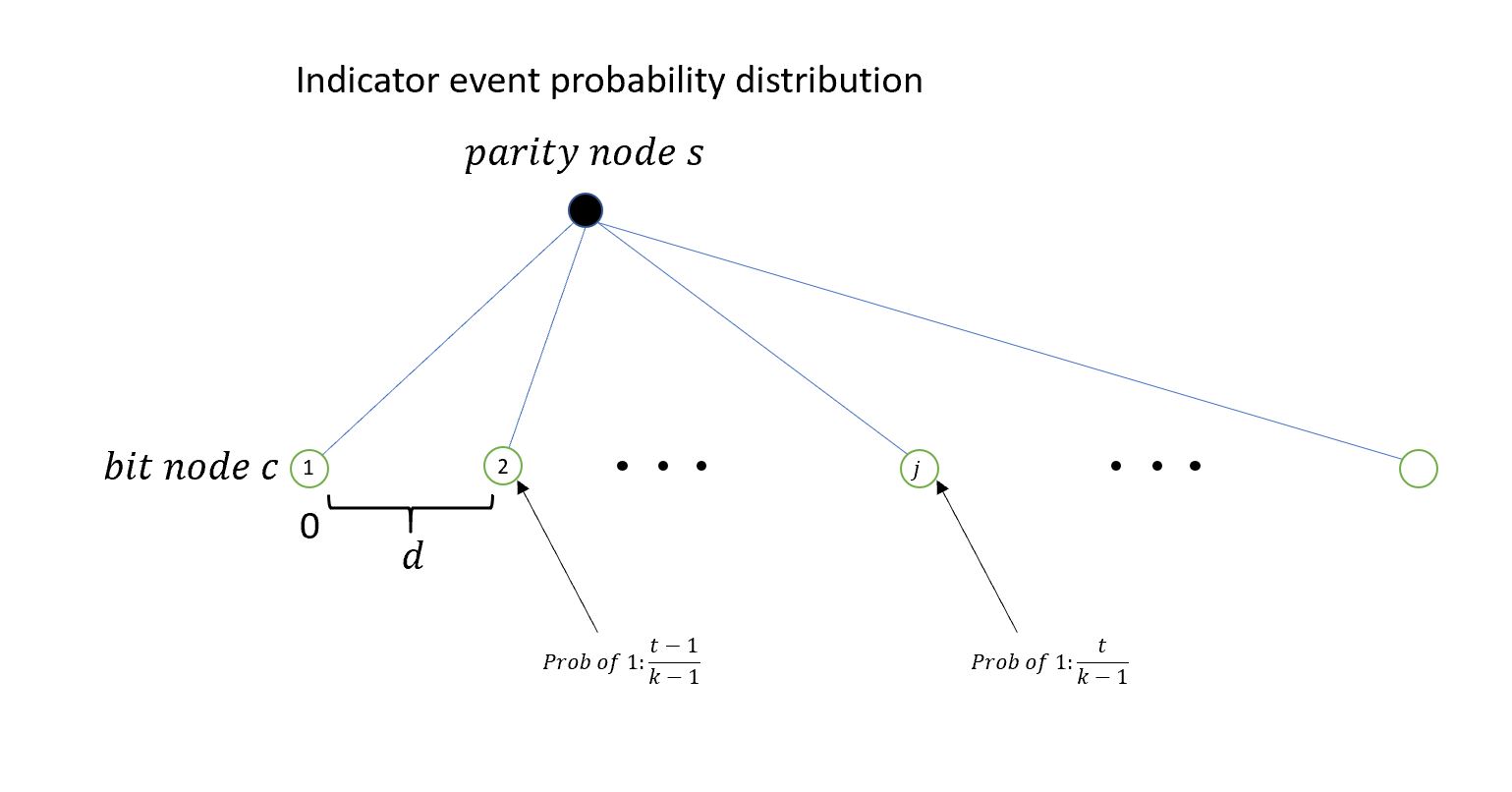}
        \caption{A figure to show the probability distribution of each bits}
        \label{fig:my_label}
    \end{figure}

    Consequently, the expectation under case 1 is:
    \begin{equation}
        \begin{aligned}\label{eq:pareto mle2}
            &\frac{1}{k-t}\sum^{k-t}_i E_{first}[\mathbf{1}^j_i,i\not\in \{a\},j=1,2|A]\\&=\frac{t-1}{k-2}\sum^{\omega_0-2}_{\mathrm{m\ is\ even}}\frac{{t-1 \choose m}{k-t-1 \choose \omega_0 -2 -m}}{{k-2 \choose \omega_0-2-m}}+\frac{k-t-1}{k-2}\sum^{\omega_0-1}_{\mathrm{m\ is\ odd}}\frac{{t \choose m}{k-t-2 \choose \omega_0-2-m}}{{k-2 \choose \omega_0-2}}
        \end{aligned}
\end{equation}

    \item Case2: $j\not=1,2$:We need to discuss the choice of position 1 and 2 first. I use a figure to show my calculation. The $Hyper(N,K,n,k$),odd/even means a hyper-geometric distribution of choosing $n$ samples with $k$ successes from $N-size$ population with $K$ successful states in population, and calculate the sum of the probability of odd/even $k$.
    
    \begin{figure}
        \centering
        \includegraphics[height=4.0cm]{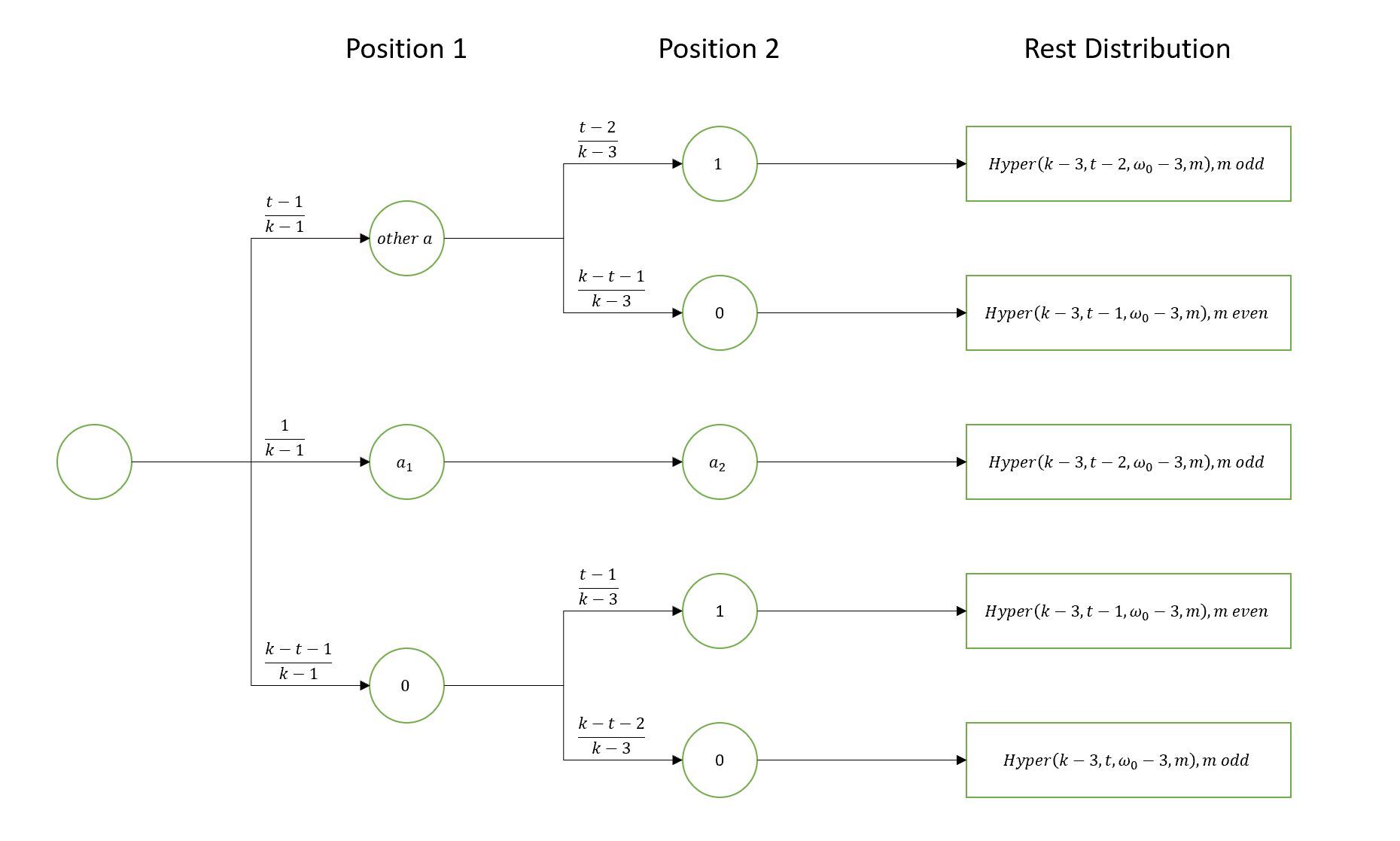}
        \caption{The process to compute the unsatisfied probability of parity-node under case2}
        \label{fig:my_label}
    \end{figure}

    Finally, the expectation under Case2 will be:
    \begin{equation}
        \begin{aligned}\label{eq:pareto mle2}
            &\frac{1}{k-t}\sum_{i}^{k-t}E_{first}[\mathbf{1}^j_i,i\not\in {a},j\not= 1,2|A]\\&=\frac{t-1}{k-1}[\frac{t-2}{k-3}\sum^{\omega_0-3}_{\mathrm{m\ is\ odd}}\frac{{t-2 \choose m}{k-t-1 \choose \omega_0-3-m}}{{k-3 \choose \omega_0-3}}+\frac{k-t-1}{k-3}\sum^{\omega_0-3}_{\mathrm{m\ is\ even}}\frac{{t-1 \choose m}{k-t-2 \choose \omega_0-3-m}}{{k-3\choose \omega_0-3}}]\\&+\frac{1}{k-1}\sum^{\omega_0-3}_{\mathrm{m\ is\ odd}}\frac{{t-2 \choose m}{k-t-2 \choose \omega_0-3-m}}{{k-3 \choose \omega_0-3}}\\&+\frac{k-t-1}{k-1}[\frac{t-1}{k-3}\sum^{\omega_0-3}_{\mathrm{m\ is\ even}}\frac{{t-1 \choose m}{k-t-2 \choose \omega_0-2-m}}{{k-3 \choose \omega_0-3}}+\frac{k-t-2}{k-3}\sum^{\omega_0-3}_{\mathrm{m\ is\ odd}}\frac{{t \choose m}{k-t-3 \choose \omega_0 -3 m}}{{k-3 \choose \omega_0 -3}}]&&
        \end{aligned}
    \end{equation}
    \item Final Expectation of Score under Event A: Case1 applied to $j=1$ or $j=2$, and Case2 applied to all other $\omega_0-2$ choices of $j$. Thus:
    \begin{equation}
        \begin{aligned}\label{eq:pareto mle2}
            E_{first}[Score_0|A] &=\frac{2}{k-t}\sum^{k-t}_iE_{first}[\mathbf{1}^j_i,i\not \in \{a\},j=1,2|A]\\&+\frac{\omega_0-2}{k-t}\sum^{k-t}_i E_{first}[\mathbf{1}^j_i,i\not \in{a},j\not=1,2|A]
        \end{aligned}
    \end{equation}
\end{itemize}
    Similarly, we assume a linear model between $Score_0$ and $\theta$. By the same analysis of Event $A$, we could get: $E_{first}[Score_0]=\beta_0E[\theta]+b_0$ and $E_{first}[Score_0|A]=\beta_0E[\theta|A]+b_0=\beta_0(E[\theta]+1)+b_0$. An estimate of $\beta_0$ can be computed by: $E_{first}[Score_0|A]-E_{first}[Score_0]$.

\section{From the first layer to the performance of decoding}
In LDPC decoding, the first layer plays a crucial role in the whole decoding process. In fact, most errors are corrected in the first three layers. To QC-MDPC code, the quasi-cyclic property ensures the symmetry among each bit. Therefore, it would be sufficient to quantify the performance of the first layer by two parameters: ErrCrt and ErrGen.

In most successful decoding, the first layer corrects a considerable amount of errors and generate very few errors. In the 90-bits security scheme, our default decoder can generally correct about 20.463 errors and cause 2.462 new errors by the default decoder.

However, the performance of decoding is not simply mostly related to ErrCrt-ErrGen, which is the net errors left after the first layer. 

To timing attack, we measure the number of iterations during decoding. The Iterations has the largest correlation to $ErrCrt-1.6ErrGen$. That’s to say, generating fewer new errors is more important than correcting existing errors in reducing decoding time.

\begin{figure}[H]
    \centering
    \includegraphics[height=4.0cm]{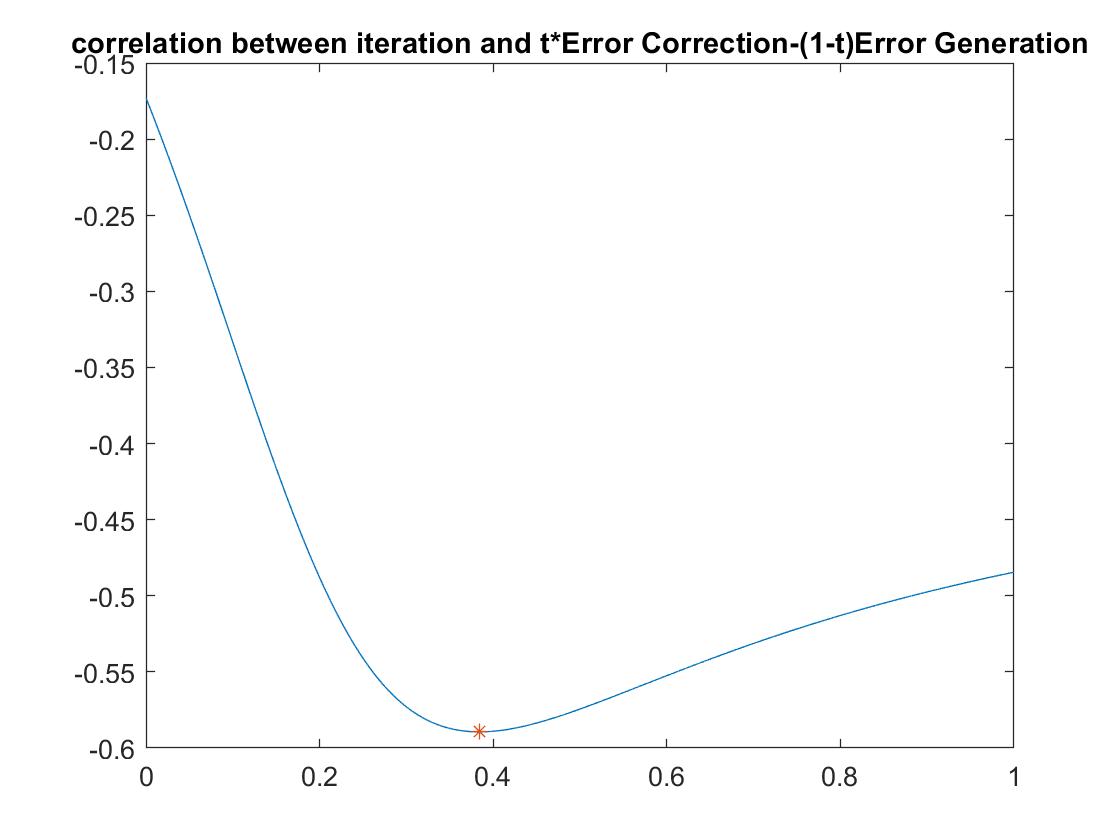}
    \caption{Correlation between timing attack and different combination of ErrCrt and ErrGen}
    \label{fig:my_label}
\end{figure}

A reasonable explanation for this relationship is that the error generating rate of the first layer is positively and closely related to the error generating rate of the following layers. In successful decoding, the new errors need two layers to complete a whole circle: one layer to generate, and one layer to eliminate. This assumption helps explain the near 1:2 ratio of Error Correction and Error Generation. Therefore, we could utilize this linear relationship to explain the timing attack.

\section{An explanation to the timing attack algorithm}\label{references}

From above analysis, we get $\delta=t_{max}ErrCrt-(1-t_{max})ErrGen$ maximizing the correlation between $\delta$ and iterations during decoding. Then, we assume a linear model between $\delta$ and iterations:

    \begin{equation}
        \begin{aligned}\label{eq:pareto mle2}
            Iteration=\beta\delta+b+\epsilon=\beta(t_{max}ErrCrt-(1-t)ErrGen)+b+\epsilon
        \end{aligned}
    \end{equation}

$\delta$, in a sense, represents the error-correcting ability of the first layer. Thus, it is believed to have a positive $\beta$. Since ErrCrt and ErrGen are linearly related to $\theta$, Iteration is also linearly related to $\theta$. 
    \begin{equation}
        \begin{aligned}\label{eq:pareto mle2}
            Iteration=\beta'\theta+b'+\epsilon'
        \end{aligned}
    \end{equation}

So, this $\beta'$ is negative since the relationship between $\theta$ and $\delta$ is negative. During the timing attack, attackers have access to the spectrum of error, $\Delta(e)$, and iterations to decode this error. Assuming that we are attacking the first position of secrete key $\Delta(h_0)_1$. Recall the definition of $\theta$.

\begin{equation}
        \begin{aligned}\label{eq:pareto mle2}
            Iteration &=\beta'(\Delta(e)_1\Delta(h_0)_1+\sum_{rest}\Delta(e)_i\Delta(h_0)_i)+b+\epsilon'\\&=\beta'\Delta(h_0)_1\Delta(e)_1+\beta'\sum_{rest}\Delta(e)_i\Delta(h_0)_i+b+\epsilon'
        \end{aligned}
\end{equation}

In the timing attack algorithm on distance 1, the program will record the number of errors whose first distance is not zeros. Mathematically, the program will return: $$\sum^{MaxTurn}_{i=1}\mathbf{1}_{\Delta(e_i)_1}$$

The $\mathbf{1}_{\Delta(e_i)_1}$ is the indicator function of $\Delta(e_i)_1$, the first term in the distance spectrum of the $i$-th error.

Also, the program will sum up the number of iterations in decoding errors which has a nonzero first term in their distance spectrum. In math, this quantity is:

\begin{equation}
        \begin{aligned}\label{eq:pareto mle2}
            \sum^{MaxTurn}_{i=1}Iteration_i\mathbf{1}_{\Delta(e_i)_1}=\sum^{MaxTurn}_{i=1}[\beta'\Delta(h_0)_1\Delta(e_i)_1\mathbf{1}_{\Delta(e_i)_1}\\+\beta'\sum_{rest}\Delta(e_i)_j\Delta(h_0)_j\mathbf{1}_{\Delta(e_i)_1}+b\mathbf{1}_{\Delta(e_i)_1}+\epsilon\mathbf{1}_{\Delta(e_i)_1}]
        \end{aligned}
\end{equation}

We evaluate the first term when $MaxTurn\xrightarrow{}\infty$: $$\sum^{MaxTurn}_{i=1}\beta'\Delta(h_0)_1\Delta(e_i)_1\mathbf{1}_{\Delta(e_i)_1}\xrightarrow{}\beta'\Delta(h_0)_1\mathbf{E}[\Delta(e)_1|\Delta(e)_1\neq0]\sum^{MaxTurn}_{i=1}\mathbf{1}_{\Delta(e_i)_1}$$

When $MaxTurn\xrightarrow{}\infty$, we expect $\sum^{MaxTurn}_{i=1}\mathbf{1}_{\Delta(e_i)_1}\xrightarrow{}\infty$. Therefore, by our assumption of independence among $\Delta(e)_j$, we can evaluate the second term to:

$$\sum^{MaxTurn}_{i=1}\beta'\sum_{rest}\Delta(e_i)_j\Delta(h_0)_j\mathbf{1}_{\Delta(e_i)_1}\xrightarrow{}\beta'\sum_{rest}\mathbf{E}[\Delta(e)_j]\Delta(h_0)_j\sum^{MaxTurn}_{i=1}\mathbf{1}_{\Delta(e_i)_1}$$

The rest two terms will also converge to:
$$\sum^{MaxTurn}_{i=1}(b+\epsilon)\mathbf{1}_{\Delta(e_i)_1}\xrightarrow{}b\sum^{MaxTurn}_{i=1}\mathbf{1}_{\Delta(e_i)_1}$$

Together, the program output is:

\begin{equation}
        \begin{aligned}\label{eq:pareto mle2}
            &\frac{\sum^{MaxTurn}_{i=1}Iteration_i\mathbf{1}_{\Delta(e_i)_1}}{\sum^{MaxTurn}_{i=1}\mathbf{1}_{\Delta(e_i)_1}}\xrightarrow{}\beta'\Delta(h_0)_1\mathbf{E}[\Delta(e)_1|\Delta(e)_1\neq0]+\beta'\sum_{rest}\mathbf{E}[\Delta(e)_j]\Delta(h_0)_j+b\\&=\beta'(\mathbf{E}[\Delta(e)_1|\Delta(e)_1\neq0]-\mathbf{E}[\Delta(e)_1])\Delta(h_0)_1+\beta'\sum_{all}\mathbf{E}[\Delta(e)_j]\Delta(h_0)_j+b
        \end{aligned}
\end{equation}

Also, the relationship among $\Delta(h_0)$ is:
$$\sum_{all}\Delta(h_0)_j={\omega_0 \choose 2}$$
Therefore, the program output will cluster into several levels which totally depends on the $\Delta(h_0)_1$ and the gap is: $$\beta'(\mathbf{E}[\Delta(h_0)_1|\Delta(h_0)_1\neq0]-\mathbf{E}[\Delta(h_0)_j])$$

\newpage
\section*{Appendix A: Some Important Algorithms}

\begin{algorithm}[H]
\caption{QC-MDPC Decoding}
\hspace*{\algorithmicindent} \textbf{Input} Codeword $c$, Parity-check matrix $H$, Threshold $b$, Max Iteration $M$\\
\hspace*{\algorithmicindent} \textbf{Output} Purified Codeword $m$ or failure symbol $\perp$
\begin{algorithmic}[1]
\STATE $s\leftarrow H*c' $ in $\mathbb{F}_2$
\STATE $i\leftarrow0$
\WHILE{$(s \neq \mathbf{0})\wedge(i<M)$}
\STATE $i\leftarrow i+1$
\STATE $c\leftarrow c+(counter>b[i])$ in $\mathbb{F}_2$
\ENDWHILE
\IF{s=$\mathbf{0}$}
\RETURN $m\leftarrow c$
\ELSE
\RETURN $\perp$
\ENDIF
\end{algorithmic}
\end{algorithm}

\begin{algorithm}[H]
\caption{Timing Attack}
\hspace*{\algorithmicindent} \textbf{Output} Purified Codeword $m$ or failure symbol $\perp$
\begin{algorithmic}[1]
\STATE Generates an error $e=[e_0|e_1],e_0,e_1\in \mathbb{F}^k_2$, where $e_1=\mathbf{0},wt(e_0)=t$
\STATE Compute the spectrum of the error $\Delta(e)$
\STATE Send $e$ to target (Here the codeword $c=\mathbf{0}$)
\STATE $n\leftarrow$ number of iterations (from side channel)
\FOR{each $i$, $\Delta(e)_i\neq0$}
\STATE $observed_i+=1$
\STATE $iteration_i+=n$
\ENDFOR
\RETURN $iteration_i/observed_i$ for $i\in \{1,2,...,\left\lfloor\frac{k}{2}\right\rfloor\}$

\end{algorithmic}
\end{algorithm}

\newpage
\section*{Appendix B: Experiment results for section 3.1}

We do experiments on 90-bit security crypto-system scheme with decoder threshold [30,28,26,25,23,23,23,23,23,23,23,23,23,23,23,23,23]. A random secret key $[h_0,h_1]$ with $ω_0=ω_1$ is generated and 50000 errors are tested on this key. In 90-bit security system, the estimate slope $\beta_1$ is:$-0.001153$.

\begin{figure}[H]
\centering
\includegraphics[height=4.0cm]{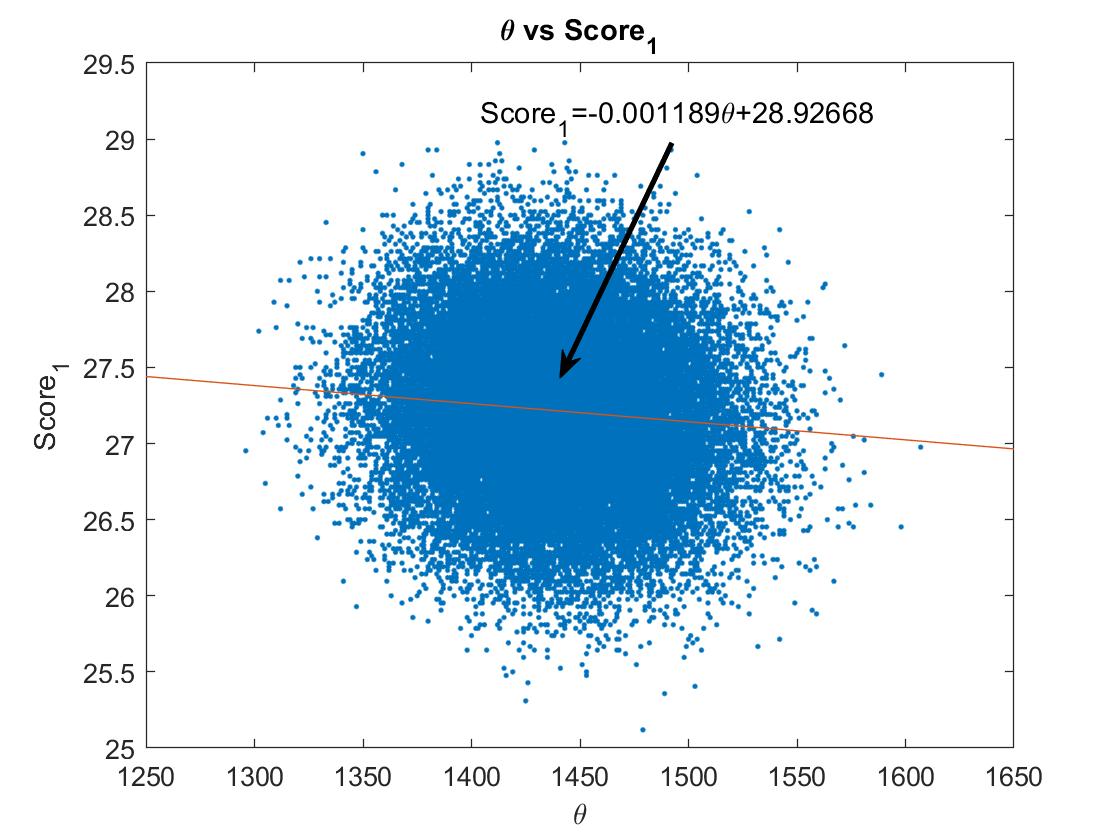}
\caption{a plot of $theta$ vs $Score_1$ based on 50000 trials on a 90-bit system and a pre-determined decoder. The red line is the regression line.}
\label{fig:example}
\end{figure}

We can see the actual slope given by this experiment is $-0.001159$. This value is very close to our estimate value.

The first layer will correct an error if it has a counter value larger than (equal to) the threshold, which is 30 in this case. The counter value of an erroneous bit does not has a notable distribution. However, since the threshold is close to its expectation, the relation between the expectation and the error correction in first layer is relatively strong.

\begin{figure}[H]
\centering
\includegraphics[height=4.0cm]{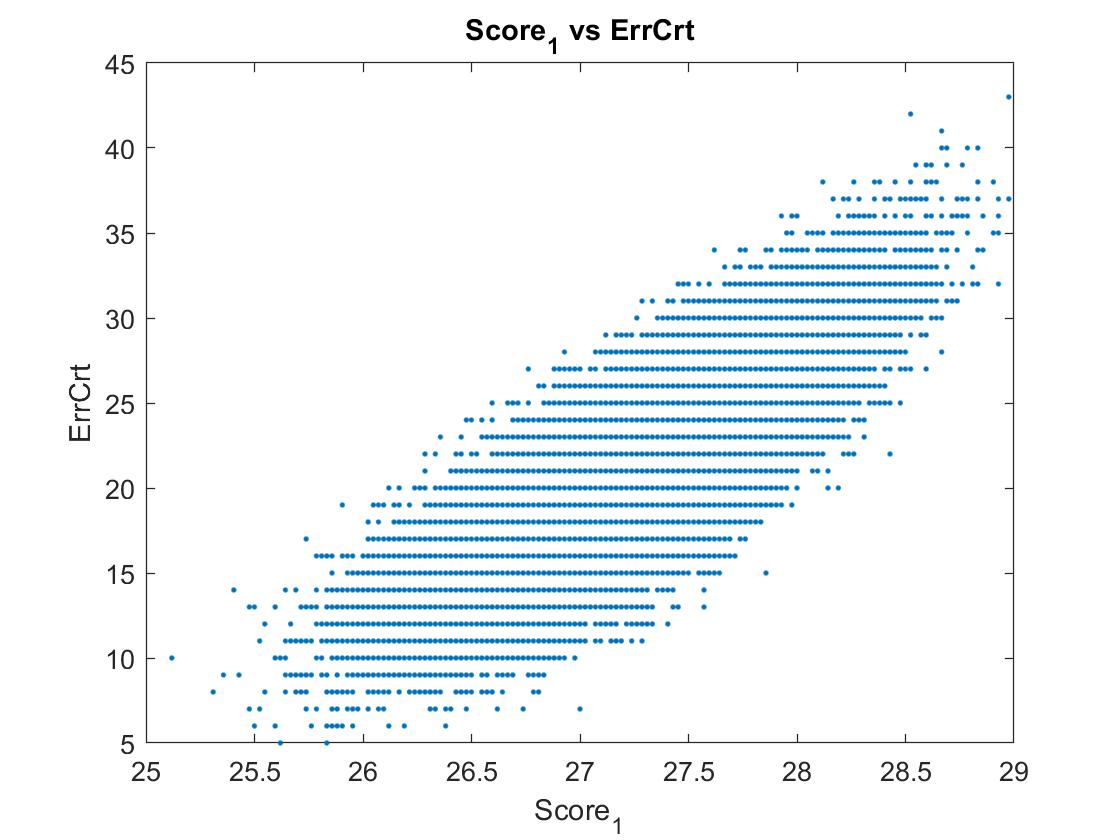}
\caption{a plot of $Score_1$ vs ErrCrt based on 50000 trials on a 90-bit system and a pre-determined decoder. An obvious linear relation can be observed.}
\label{fig:example}
\end{figure}

\newpage

\section*{Appendix C: Applying the result in section 3.2 to second half part of the codeword }

All bits in the second half part are errorless, thus has no influence on the parity nodes. Also, we have no information on the secret key $h_1$. Thus, bits in the second half part is randomly connected to $\omega_1$ parity nodes. All parity nodes are symmetric thus has a same probability to be unsatisfied (but not necessarily independent). Then,

\begin{equation}
    \begin{aligned}\label{eq:pareto mle2}
        E_{second}[Score_0]=\omega_1Pr(s=1)=\omega_1\frac{wt(s)}{k}
    \end{aligned}
\end{equation}
    
every unsatisfied parity node will increase the total score in the first half by exactly $\omega_0$. So:

\begin{equation}
    \begin{aligned}\label{eq:pareto mle2}
        (k-t)E_{first}[Score_0]+tE[Score_1]=\omega_0wt(s)
    \end{aligned}
\end{equation}
    
The crypto-system requires $k-t>>t$:

\begin{equation}
    \begin{aligned}\label{eq:pareto mle2}
        wt(s)\approx\frac{k-t}{\omega_0}E_{first}[Score_0]
    \end{aligned}
\end{equation}
    
Thus:
    
\begin{equation}
    \begin{aligned}\label{eq:pareto mle2}
        E_{second}[Score_0]\approx\frac{\omega_1(k-t)}{\omega_0k}E_{first}[Score_0]
    \end{aligned}
\end{equation}
    
A good QC-MDPC crypto-system will require $\omega_0\approx\omega_1$. Otherwise, the decoder of this QC-MDPC code will be inefficient. Therefore,

\begin{equation}
    \begin{aligned}\label{eq:pareto mle2}
        E[Score_0]\approx E_{second}[Score_0]\approx E_{first}[Score_0]
    \end{aligned}
\end{equation}
    
The conclusion on $E_{first}[Score_0]$ can be applied to $E[Score_0]$.

\newpage
\section*{Appendix D: Experiment results for section 3.2}
From the above analysis, the slope between $Score_0$ and $\theta$ for 90-bit security is predicted to be -0.003979. The experiment result described in the previous section gives us a slope of -0.004172.

\begin{figure}[H]
    \centering
    \includegraphics[height=6.2cm]{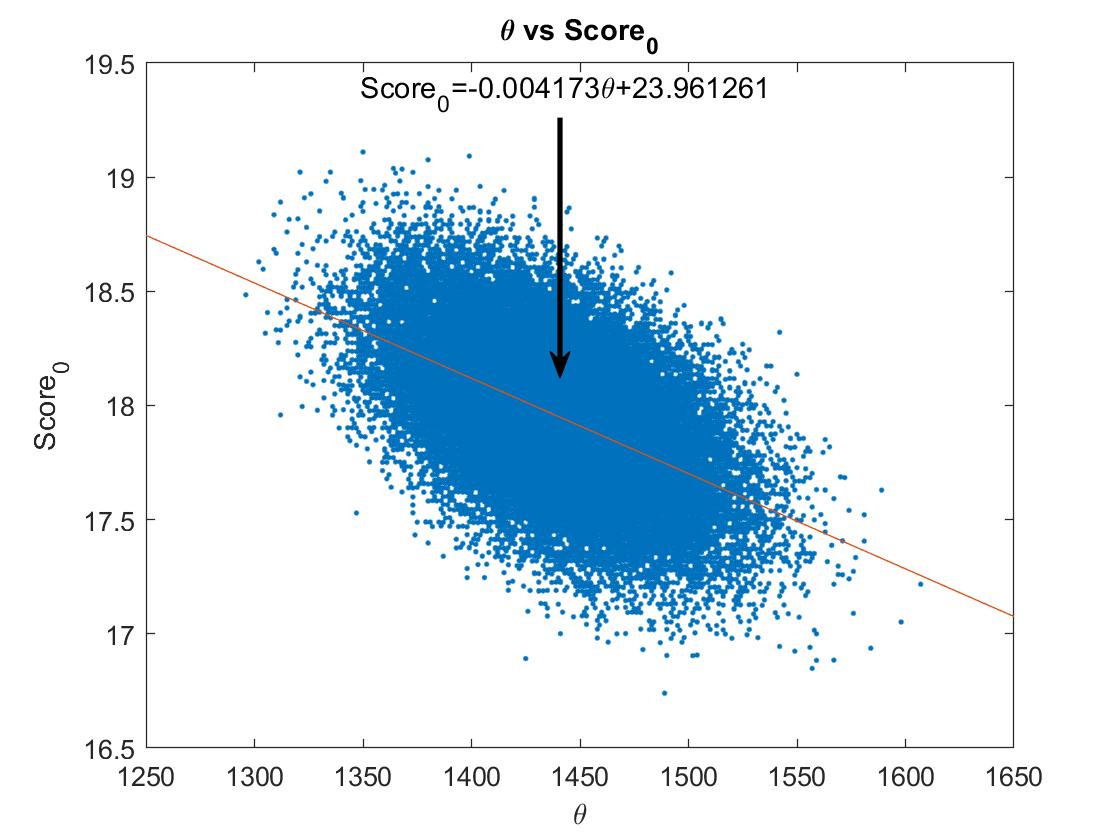}
    \caption{The linear relationship between $\theta$ and the average $Score_0$}
    \label{fig:my_label}
\end{figure}

When we have a close look up of the distribution of $Score_0$. It generally follows a normal distribution.

\begin{figure}[H]
    \centering
    \includegraphics[height=5.0cm]{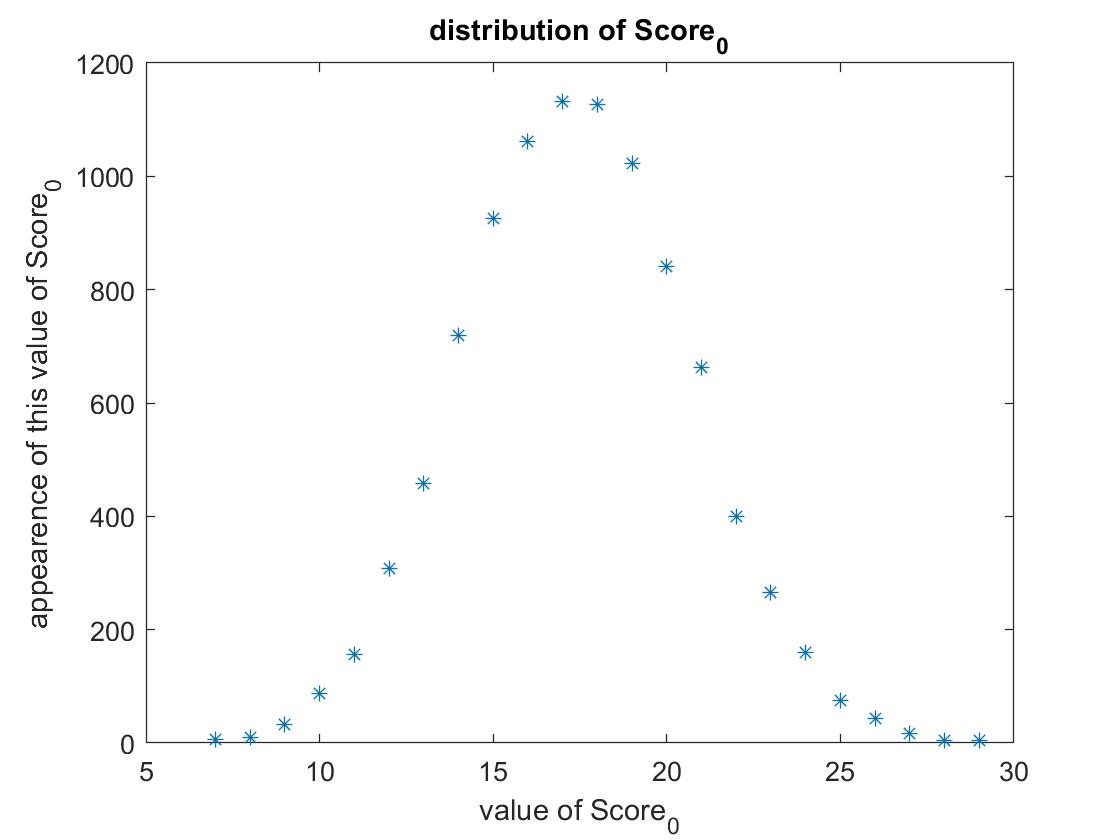}
    \caption{A figure to show the distribution of $Score_0$}
    \label{fig:my_label}
\end{figure}

New errors are generated in first layer if the score of an errorless bit exceeds the threshold of the decoder, which is 30 in our test. It is obvious that an decrease in the expectation of score will result in a decrease in the false flip, ErrGen.

\begin{figure}[H]
    \centering
    \includegraphics[height=5.0cm]{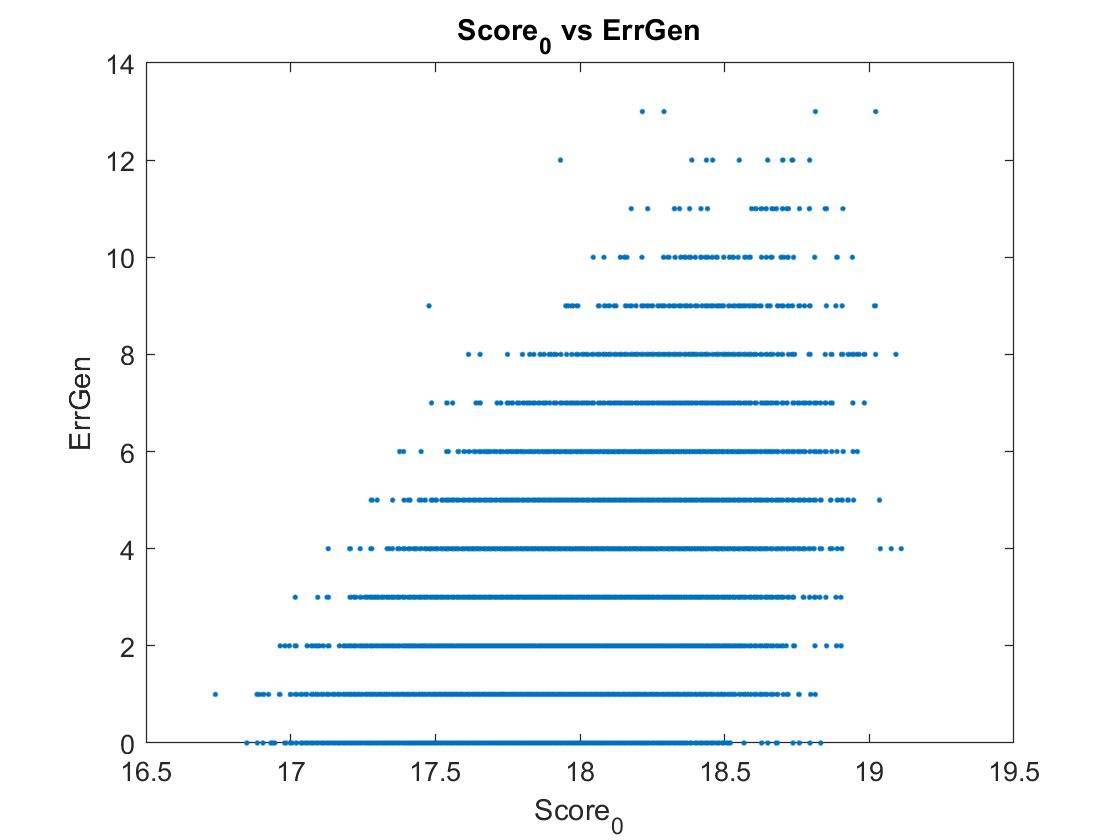}
    \caption{the plot of $Score_0$ vs ErrGen}
    \label{fig:my_label}
\end{figure}

From the above figure, These two variables have a positive relationship. Moreover, the left side has a clear linear relationship. This observation provides that a low average $Score_1$ ensures a low new error generation rate, but high expectation does not necessarily result in high error generation rate.

\newpage
\section*{Appendix E: Experiment results for section 5}
It's difficult to conduct experiment on iteration directly because the slope $\beta'$ is usually determined by experiment. But, we can test on the linear relationship between $Score_0$ and $\Delta(h_0)_1$. Similarly,
\begin{equation}
        \begin{aligned}\label{eq:pareto mle2}
            \frac{\sum^{MaxTurn}_{i=1}(Score_0)_i\mathbf{1}_{\Delta(e_i)_1}}{\sum^{MaxTurn}_{i=1}\mathbf{1}_{\Delta(e_i)_1}}\xrightarrow{}\beta_0\Delta(h_0)_1\mathbf{E}[\Delta(e)_1|\Delta(e)_1\neq0]+\beta_0\sum_{rest}\mathbf{E}[\Delta(e)_j]\Delta(h_0)_j+b
        \end{aligned}
\end{equation}
As discussed in section 3, on 90-bits security protocol, the value of $\beta_0$ is predicted to be $-0.003979$. The value of $(\mathbf{E}[\Delta(h_0)_1|\Delta(h_0)_1\neq0]-\mathbf{E}[\Delta(h_0)_j])$ can be easily determined both experimentally and theoretical. On 90-bits protocol, it is about $0.4293$. So, the clustering gap is predicted to be $-0.003979*0.429140=-0.001708$.

\begin{figure}
\centering
\includegraphics[height=4.8cm]{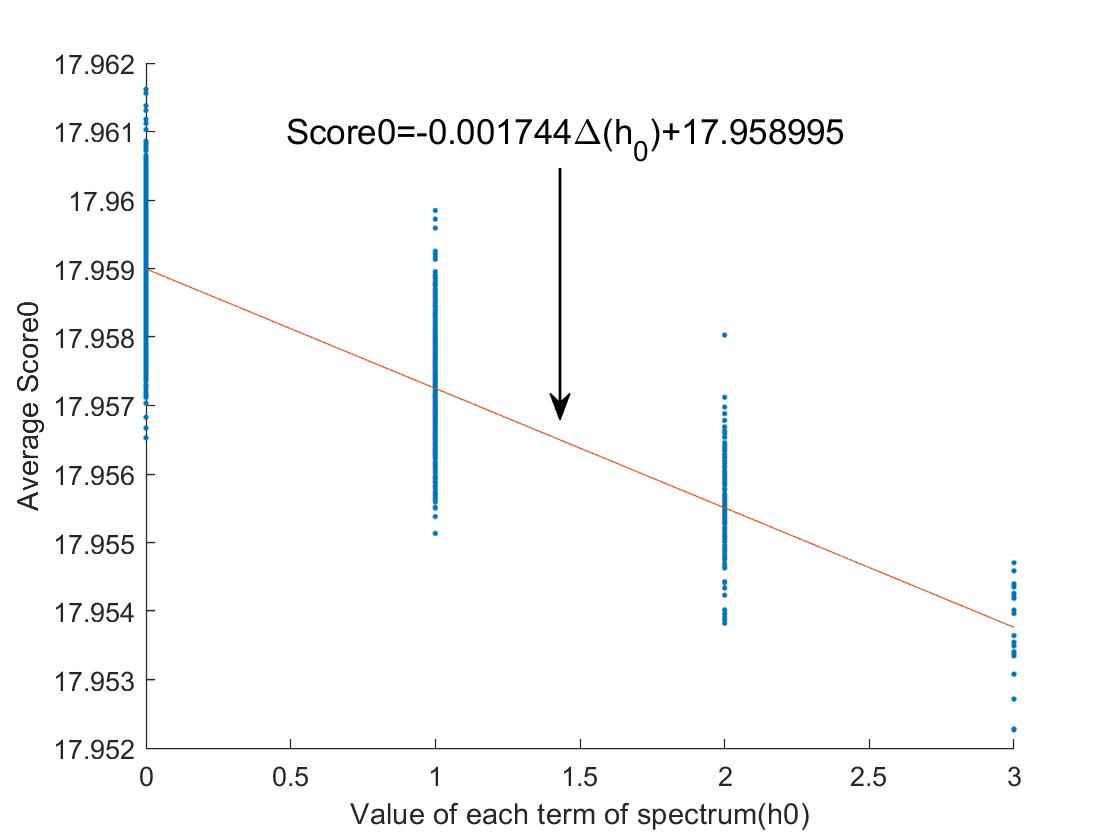}
\caption{An attack on $Score_0$ based on 50000 trials. We can see a clear linear relationship.}
\label{fig:example}
\end{figure}

Then, an attack on $Score_0$ based on $50000$ trials is performed. The gap is about $-0.001744$, which is very close to my prediction.

\end{document}